\documentclass[aps,prl,twocolumn,superscriptaddress,showpacs]{revtex4}

\usepackage{amsmath}
\usepackage{graphics}
\usepackage{natbib}
\usepackage{amssymb}
\usepackage{bm}

\begin{document}

\title{Spin and orbital Hall effects for diffracting optical beams in gradient-index media}
\author{Konstantin Y. Bliokh}
\affiliation{Nonlinear Physics Centre, Research School of Physics and Engineering, The Australian National University, Canberra ACT 0200, Australia}
\affiliation{Institute of Radio Astronomy, 4 Krasnoznamyonnaya Street, Kharkov 61002, Ukraine}
\author{Anton S. Desyatnikov}
\affiliation{Nonlinear Physics Centre, Research School of Physics and Engineering, The Australian National University, Canberra ACT 0200, Australia}

\begin{abstract}
We examine the evolution of paraxial beams carrying intrinsic spin and orbital angular momenta (AM) in gradient-index media. A parabolic-type equation is derived which describes the beam diffraction in curvilinear coordinates accompanying the central ray. The center of gravity of the beam experiences transverse AM-dependent deflections -- the spin and orbital Hall effects. The spin Hall effect generates a transverse translation of the beam as a whole, in precise agreement with recent geometrical optics predictions. At the same time, the orbital Hall effect is significantly affected by the diffraction in the inhomogeneous medium and is accompanied by changes in the intrinsic orbital AM and deformations of the beam.

\end{abstract}

 \maketitle

\section{I. Introduction}

Spin-dependent transverse transport -- the spin-Hall effect (SHE) -- of classical waves and quantum particles is currently attracting growing attention in condensed-matter~\cite{SHE}, high-energy~\cite{Dirac}, and optical~\cite{LZ,Bliokh1,OMN,DHH,HK,NP} physics. This effect appears under bending of the wave trajectory in an external potential and is closely related to
such fundamental phenomena as
the Berry phase, conservation of the total angular momentum (AM) of the wave, and spin-orbit interaction.

The optical SHE deals with evolution of Gaussian-type wave beams bearing intrinsic \emph{spin} AM. A generalization of this effect for higher-order Laguerre-Gaussian-type beams with phase singularities (vortices) has been put forward recently~\cite{Bliokh2,Fedo,OS}. As such beams carry well-defined intrinsic \emph{orbital} AM~\cite{OAM}, an orbital-Hall effect (OHE) appears under the bending of their trajectories.

Extensive studies over the past several years mostly considered the semiclassical trajectory equations tracing evolution of the \emph{center of gravity} of a wave beam rather than the propagation of real \emph{extended} beams.
At the same time, the typical transverse shift of the beam's center of gravity due to the SHE or OHE is proportional to the wavelength and is rather small as compared to the characteristic scale of the beam deformations in an inhomogeneous medium. Therefore, it is important to give a picture of the evolution of realistic beams in a gradient-index medium, which includes SHE, OHE, and the diffraction processes.

Below we provide such a description of paraxial optical beams carrying spin and orbital AM and evolving in a smooth gradient-index medium. Although we consider Maxwell equations, our analysis can readily be extended to quantum wave equations describing evolution of quantum particles in external potentials. In particular, the OHE arises from the Laplace operator in curvilinear coordinates and is universal for any beams with vortices.
\section{II. Maxwell equations in the ray-accompanying coordinate frame}

Maxwell equations for the monochromatic electric field ${\bf E}$ in a gradient-index dielectric medium read
\begin{equation}
\left( {k_0^{-2} {\bm \nabla}^2  + \varepsilon } \right){\bf E} - k_0^{-2} {\bm \nabla} \left( {{\bm\nabla}  \cdot {\bf E}} \right) = 0,
\label{1}
\end{equation}
where $k_0  = \omega/c$ ($\omega$ is the wave frequency) and $\varepsilon  = \varepsilon \left( {\bf r} \right)$ is the dielectric constant of the medium. Classical geometrical optics (GO) shows that in the short-wavelength limit the wave propagates as a classical point particle moving along the ray trajectory ${\bf r}_c  = {\bf r}_c \left( s \right)$ given by~\cite{KO}
\begin{equation}
{\bf \dot r}_c  = {\bf t}~,~~{\bf \dot t} = \frac{{{\bm\nabla}_ \bot  \varepsilon _c }}{{2\varepsilon _c }}.
\label{2}
\end{equation}
Here the overdot stands for the derivative with respect to the parameter $s$, which is the trajectory arc length, ${\bf t}$ is the unit vector tangent to the trajectory, ${\bm\nabla}_ \bot   = {\bm\nabla} - {\bf t}\left( {{\bf t} \cdot {\bm\nabla} } \right)$ is the gradient in the plane orthogonal to the ray, and the subscript ``c'' means that the function is taken on the trajectory, i.e., ${\bm\nabla}_\bot  \varepsilon _c \equiv \left. {\left( {{\bm\nabla}_ \bot  \varepsilon } \right)} \right|_{{\bf r} = {\bf r}_c}$, etc. Equations~(\ref{2}) define the central reference ray and realistic beams evolve in the vicinity of it.

The wave-beam propagation is described using paraxial approximation in the vicinity of the GO ray~(\ref{2}). This implies the smallness of the two parameters:
\begin{equation}
\mu _1  = \frac{\lambda }{w} \ll 1 \;\;\; \text{and} \;\;\; \mu _2  = \frac{w}{L} \ll 1,
\label{3}
\end{equation}
where $\lambda$ is the wavelength, $w$ is the characteristic beam width, and $L \sim \left| {\bm\nabla} \varepsilon / \varepsilon  \right| ^{-1}$ is the characteristic scale of the medium inhomogeneity. We aim to describe the evolution of a paraxial beam keeping the terms up to the $\mu^3$ order with respect to the combined parameter $\mu  = \max \left( {\mu _1 ,\mu _2 } \right)$ in Maxwell equations. Previously, this problem has been solved with an accuracy of $\mu^2$~\cite{PS}, i.e., in the lowest-order approximation that describes diffraction but does not account for the Hall effects.
\begin{figure}[tbh]
\centering \scalebox{0.75}{\includegraphics{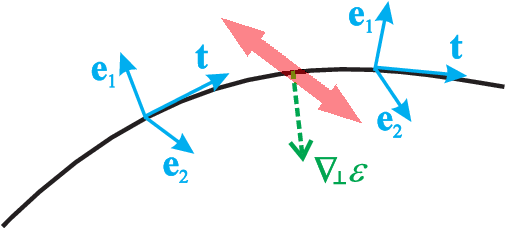}} \caption{(Color online.) Geometry of the wave propagation along a curved geometrical-optics ray. Depicted are: a ray accompanying coordinate frame attached to the unit vectors $({\bf e}_1,{\bf e}_2,{\bf t})$, direction of the inhomogeneity bending the ray, ${\bm \nabla}_{\bot} \varepsilon_c \parallel {\bf \dot{t}}$, and directions of the Hall effects orthogonal both to ${\bf t}$ and ${\bf \dot {t}}$ (the large double arrow).} \label{Fig1}
\end{figure}

In the vicinity of the GO ray, we introduce a ray-accompanying coordinate frame $\left( {\xi _1 ,\xi _2 ,s} \right)$ attached to the unit vectors $\left( {{\bf e}_1 ,{\bf e}_2 ,{\bf t}} \right)$. These vectors evolve along the ray: ${\bf t} = {\bf t}\left( s \right)$, ${\bf e}_i  = {\bf e}_i \left( s \right)$, $i=1,2$, see Fig.~\ref{Fig1}. To make the $\left( {\xi _1 ,\xi _2 ,s} \right)$ frame orthogonal, $\left( {\bf e}_1 ,{\bf e}_2  \right)$ must obey the parallel transport equation along the ray~\cite{PS,tubes}:
\begin{equation}
{\bf \dot e}_i  =  - \left( {{\bf e}_i  \cdot {\bf \dot t}} \right){\bf t},
\label{4}
\end{equation}
The Lame coefficients of the coordinates $\left( {\xi _1 ,\xi _2 ,s} \right)$ are equal to~\cite{PS,tubes}
\begin{equation}
h_1  = h_2  = 1~,~~h_s  \equiv h = 1 - \frac{{{\bm\nabla}_ \bot  \varepsilon _c }}{{2\varepsilon _c }} \cdot \bm{\xi},
\label{5}
\end{equation}
where ${\bm \xi}$ is the radius-vector in the $\left( {\xi _1 ,\xi _2 } \right)$ plane.

The electromagnetic wave is near-transverse in the ray coordinates:
\begin{equation}
{\bf E} = {\bf E}_ \bot   + E_\parallel  {\bf t} = E_i {\bf e}_i  + E_\parallel  {\bf t}~,~~
\left| {E_\parallel  } \right| \sim \mu  \left| E \right|.
\label{6}
\end{equation}
From the equation ${\bm\nabla} \cdot \left( {\varepsilon {\bf E}} \right) = 0$, stemming from Eq.~(\ref{1}), it follows that in the lowest-order approximation in $\mu$,
\begin{equation}
{\bm\nabla} \cdot {\bf E} \simeq  - \frac{{{\bm\nabla}_\bot  \varepsilon _c  }}{{\varepsilon _c  }} \cdot {\bf E}_\bot~~{\rm and}~~ E_\parallel   \simeq ik_c ^{ - 1} {\bm\nabla} _ \bot   \cdot {\bf E}_ \bot ,
\label{7}
\end{equation}
where $k_c = k_0 \sqrt {\varepsilon _c }$ is the central wave number.

Using Eq.~(\ref{5}) and the first of Eqs.~(\ref{7}), Maxwell Eq.~(\ref{1}) in the $\left( {\xi _1 ,\xi _2 ,s} \right)$ coordinates takes the form
\begin{eqnarray}
\nonumber
k_0^{-2} \left[ {\frac{1}{h}\frac{\partial }{{\partial s}}\left( {\frac{1}{h}\frac{{\partial {\bf E}}}{{\partial s}}} \right) + \frac{1}{h}\frac{\partial }{{\partial \xi _i }}\left( {h\frac{{\partial {\bf E}}}{{\partial \xi _i }}} \right)} \right]+ \varepsilon {\bf E}  \\
+ k_0^{-2} {\bm\nabla} \left( {\frac{{{\bm\nabla}_\bot  \varepsilon _c  }}{{\varepsilon _c  }} \cdot {\bf E}_ \bot  } \right) = 0.
\label{8}
\end{eqnarray}
Equation~(\ref{8}) can be projected onto the plane $\left( {\xi _1 ,\xi _2 } \right)$ orthogonal to the ray. In so doing, we notice that~\cite{remark1}
\begin{eqnarray}
\left( {\frac{{\partial {\bf E}}}{{\partial s}}} \right)_ \bot = \left( \frac{{\partial {\bf E}_\bot}}{{\partial s}} \right)_\bot + E_\parallel  {\bf \dot t},~ E_\parallel  {\bf \dot t}= i \left( {{\bm\nabla}_\bot \cdot {\bf E}_\bot  } \right)\frac{{{\bm\nabla}_\bot  \varepsilon _c }}{2\varepsilon_c k_c},
\label{9}
\end{eqnarray}
where we used Eqs.~(\ref{2}), (\ref{6}) and~(\ref{7}). This yields the wave equation for the transverse electric field ${\bf E}_ \bot$:
\begin{eqnarray}
\nonumber
k_0^{-2} \left[ {\frac{1}{h}\frac{\partial }{{\partial s}}\left( {\frac{1}{h}\frac{{\partial {\bf E}_ \bot  }}{{\partial s}}} \right) + \frac{1}{h}\frac{\partial }{{\partial \xi _i }}\left( {h\frac{{\partial {\bf E}_ \bot  }}{{\partial \xi _i }}} \right)} \right]_\bot + \varepsilon {\bf E}_ \bot  \\ + k_0^{-2} \left( {\frac{{{\bm\nabla}_\bot  \varepsilon _c  }}{{\varepsilon _c  }} \times {\bm\nabla}_\bot  } \right) \times {\bf E}_ \bot   = 0.
\label{10}
\end{eqnarray}

As we will see, the last term in Eq.~(\ref{10}), which is of the order of $\mu^3$, describes the SHE of light. This term originates from the combination of the polarization term ${\bm\nabla}_\bot  \left( {{\bm\nabla} \cdot {\bf E}} \right)$ and the Coriolis term $2ik_c  E_\parallel  {\bf \dot t}$~\cite{remark1}.

\section{III. Parabolic-type equation}
Full wave Eq.~(\ref{10}) can further be simplified to a parabolic-type equation via the WKB \textit{ansatz}:
\begin{eqnarray}
{\bf E}_ \bot \left( {s,{\bm \xi }} \right) = {\bf e}\left( s \right)\varepsilon _c ^{ - 1/4} \left( s \right)W\left( {s,{\bm \xi }} \right) e^{i\Phi(s)},
\label{11}
\end{eqnarray}
where $\Phi(s)={k_0 \int\limits_0^s {\sqrt {\varepsilon _c (s') } } ds'}$ is the GO phase along the ray, ${\bf e}\left( s \right) = \alpha {\bf e}_1 \left( s \right) + \beta {\bf e}_2 \left( s \right)$ is the unit polarization vector, ${\bf e}^*  \cdot {\bf e} = 1$, and $W\left( {s,{\bm \xi }} \right)$ is the unknown slowly varying envelope of the wave. The components of the polarization vector is unchanged along the ray, $\alpha ,\beta  = {\mathop{\rm const}\nolimits}$, which signifies the parallel-transport law for the wave electric field, related to the Berry phase~\cite{PS,GP}. Owing to Eq.~(\ref{4}), derivatives ${\bf \dot e}_i \parallel {\bf t}$ do not contribute to the equation (\ref{10}) for ${\bf E}_ \bot$. Note that only the last term in Eq.~(\ref{10}) involves the wave polarization, and this term is diagonalized in the basis of circular polarizations: ${\bf e} = {\bf e}_1  + i\sigma {\bf e}_2$, where $\sigma  =  \pm 1$ is the wave helicity due to the spin. Substituting Eq.~(\ref{11}) with a circular polarization into Eq.~(\ref{10}) and retaining only terms up to the $\mu^3$ order, we arrive at the parabolic-type equation for $W$:
\begin{widetext}
\begin{eqnarray}
\nonumber
2ik_c  \frac{{\partial W}}{{\partial s}} + \Delta _\bot  W + k_0^{2} \left[ {\frac{1}{2}\left( {{\bm \xi } \cdot {\bm\nabla}_\bot  } \right)^2 \varepsilon _c   - \frac{3}{4 \varepsilon _c}\left( {{\bm \xi } \cdot {\bm\nabla}_\bot  } \varepsilon _c \right)^2   } \right]W = i\sigma \left( {\frac{{{\bm\nabla}_\bot  \varepsilon _c  }}{{\varepsilon _c  }} \times {\bm\nabla}_\bot  W} \right) \cdot {\bf t} + \frac{{{\bm\nabla}_\bot  \varepsilon _c  }}{{\varepsilon _c  }} \cdot {\bm\nabla}_ \bot  W \\
- 2ik_c  \left( {{\bm \xi } \cdot \frac{{{\bm\nabla}_ \bot  \varepsilon _c  }}{{\varepsilon _c  }}} \right)\frac{{\partial W}}{{\partial s}} - ik_c  \left( {{\bm \xi } \cdot \frac{\partial }{{\partial s}}\frac{{{\bm\nabla}_ \bot  \varepsilon _c  }}{{2\varepsilon _c  }}} \right)W - \frac{1}{6} k_0^{2} \left[ {\left( {{\bm \xi } \cdot {\bm\nabla}_\bot  } \right)^3 \varepsilon _c  } \right]W.
\label{12}
\end{eqnarray}
\end{widetext}
Here ${\bm\nabla}_\bot   = \partial /\partial {\bm \xi }$, $\Delta _ \bot   = \frac{{\partial ^2 }}{{\partial \xi _1^2 }} + \frac{{\partial ^2 }}{{\partial \xi _2^2 }}$, and in derivation of Eq.~(\ref{12}) we used the Taylor expansion for $\varepsilon \left( {\bf r} \right)$:
\[ \varepsilon  \simeq \varepsilon _c   + \left( {{\bm \xi } \cdot {\bm\nabla}_\bot  } \right)\varepsilon _c   + \frac{1}{2}\left( {{\bm \xi } \cdot {\bm\nabla}_ \bot  } \right)^2 \varepsilon _c   + \frac{1}{6}\left( {{\bm \xi } \cdot {\bm\nabla}_ \bot  } \right)^3 \varepsilon _c.\]

The first two terms of Eq.~(\ref{12}) represent usual parabolic equation, whereas the next terms in square brackets describe the influence of the medium inhomogeneity on diffraction. All the terms in the left-hand side of Eq.~(\ref{12}) are of the order of $\mu^2$; they have been obtained in Ref.~\cite{PS}. The right-hand side of Eq.~(\ref{12}) represent corrections of the $\mu^3$ order.

As we will see, the first two terms in the right-hand side of Eq.~(\ref{12}), which are proportional to ${\bm\nabla}_\bot  W$, describe the SHE and OHE. (Note that the second term originates from the $\frac{{\partial h}}{\partial {\bm \xi }}\frac{\partial }{\partial {\bm \xi }}$ term in the Laplace operator and has a universal form for waves of any nature~\cite{remark2}.) These two terms can be eliminated by a simple transformation:
\begin{eqnarray}
W\left( {s,{\bm \xi }} \right) = \tilde W\left( {s,{\bm \xi } - {\bm \delta }\left( s \right)} \right).
\label{13}
\end{eqnarray}
Substituting Eq.~(\ref{13}) into Eq.~(\ref{12}), we notice that $\frac{{\partial W}}{{\partial s}} = \frac{{\partial \tilde W}}{{\partial s}} - {\bm{\dot \delta}}\cdot {\bm\nabla}_\bot  \tilde W$ and the above-mentioned terms are canceled when
\begin{eqnarray}
{\bm {\dot \delta} } = \frac{\sigma }{{k_c  }}\left( {\frac{{{\bm\nabla}_\bot  \varepsilon _c  }}{{2\varepsilon _c  }} \times {\bf t}} \right) + \frac{i}{{k_c  }}\frac{{{\bm\nabla}_\bot  \varepsilon _c  }}{{2\varepsilon _c  }}.
\label{14}
\end{eqnarray}
As a result, the parabolic equation acquires the form
\begin{widetext}
\begin{eqnarray}
\nonumber
2ik_c  \frac{{\partial \tilde W}}{{\partial s}} + \Delta _ \bot  \tilde W + k_0^{2} \left[ {\frac{1}{2}\left({{\bm \xi } \cdot {\bm\nabla}_\bot} \right)^2 \varepsilon _c   - \frac{3}{4 \varepsilon _c}\left( {{\bm \xi } \cdot {\bm\nabla}_\bot  } \varepsilon _c \right)^2 } \right]\tilde W = \\
- 2ik_c  \left( {{\bm \xi } \cdot \frac{{{\bm\nabla}_\bot  \varepsilon _c  }}{{\varepsilon _c  }}} \right)\frac{{\partial \tilde W}}{{\partial s}} - ik_c  \left( {{\bm \xi } \cdot \frac{\partial }{{\partial s}}\frac{{{\bm\nabla}_\bot  \varepsilon _c  }}{{2\varepsilon _c  }}} \right)\tilde W - \frac{1}{6} k_0^{2} \left[ {\left( {{\bm \xi } \cdot {\bm\nabla}_\bot  } \right)^3 \varepsilon _c  } \right]\tilde W .
\label{15}
\end{eqnarray}
\end{widetext}
\section{IV. Spin and orbital Hall effects}
\begin{figure}[t]
\centering \scalebox{0.45}{\includegraphics{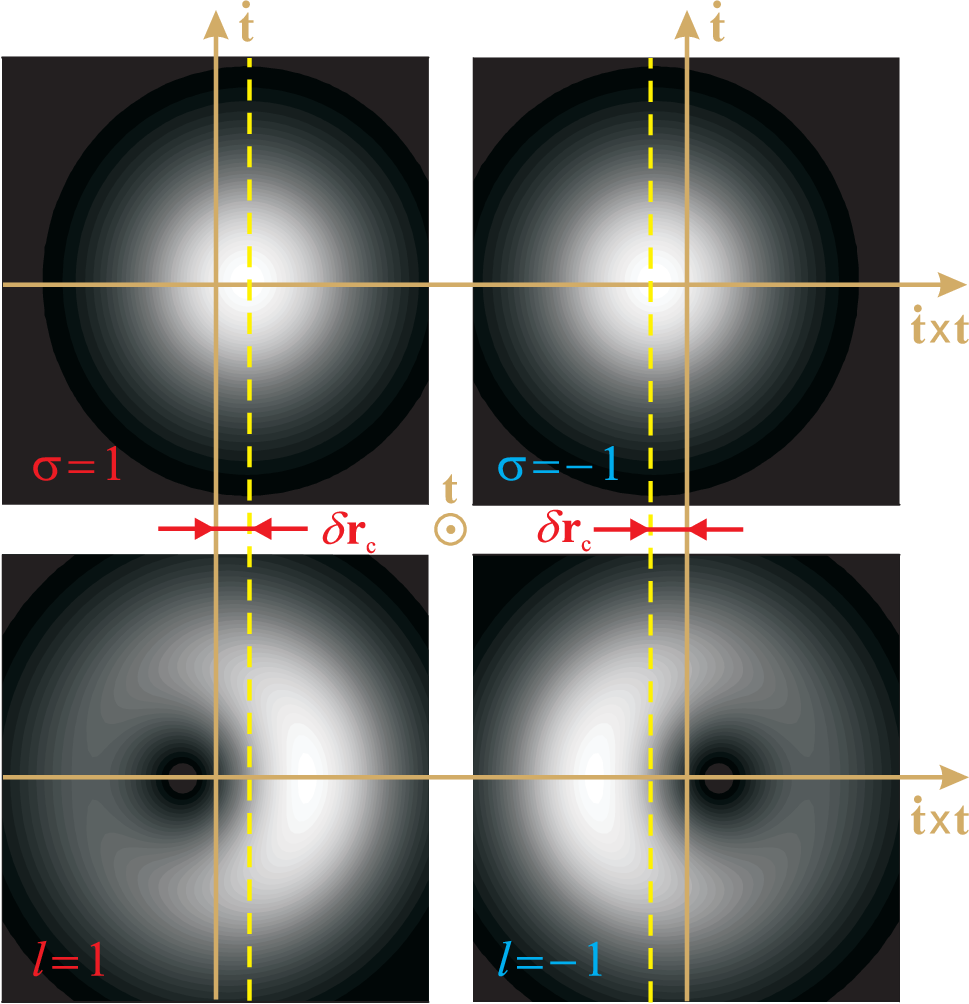}} \caption{(Color online.) Schematic picture of the SHE and OHE transverse deformations of the beam with respect to the classical GO ray, Eq.~(\ref{2}). Shown are the cases of of circularly polarized Gaussian beams with $\sigma = \pm 1$ and $l=0$ (upper panels) and of vortex Laguerre-Gaussian beams with $l = \pm 1$ and $\sigma=0$ (lower panels). The intensity distributions are plotted in arbitrary units using transformation~(\ref{13}) with some ${\bm \delta}\propto \sigma {{\bf \dot t} \times {\bf t}} +i {\bf \dot t}$, Eq.~(\ref{14}).} \label{Fig2}
\end{figure}

Equations~(\ref{13})--(\ref{15}) are the central results of the paper. While Eqs.~(\ref{13}) and (\ref{14}) describe the transverse deformations of the beam due to the SHE and OHE, Eq.~(\ref{15}) describe all other deformations caused by the diffraction in the medium. The $\sigma$-dependent term in Eq.~(\ref{14}) is responsible for SHE; it takes the form as predicted by the modified GO theories~\cite{LZ,Bliokh1,OMN,DHH,NP}. Equations (\ref{13}) and (\ref{15}) (which is independent of polarization) imply that \emph{the SHE produces a perfect translation of the whole beam} in the transverse direction, without any other polarization-dependent distortions, see Fig.~\ref{Fig2}.

The OHE is more intricate and is described by the \textit{imaginary} term in Eq.~(\ref{14}). To show this, let us first consider an unperturbed beam carrying a well-defined intrinsic orbital AM. Its field contains an \emph{optical vortex}, which is described by the structure~\cite{OAM}
\begin{eqnarray}
W_l  \propto \left[ {\xi _1  + i\,{\mathop{\rm sign}} (l)\,\xi _2 } \right]^{\left| l \right|}  = \rho ^{\left| l \right|} \exp \left( {il\varphi } \right),
\label{16}
\end{eqnarray}
where $l = 0, \pm 1, \pm 2,...$ is the vorticity, characterizing the value of intrinsic orbital AM per photon, whereas $\left( {\rho ,\varphi } \right)$ are the polar coordinates in the ${\bm \xi}$ plane. It is easy to see that small imaginary shift along, say, the $\xi_1$ axis: $\xi _1  \to \xi _1  - i\chi$, deforms the intensity distribution in the vortex along the orthogonal $\xi_2$ axis:
\[
\left| {W_l } \right|^2  \propto \left[ {\xi _1^2  + \left( {\xi _2  - {\mathop{\rm sign}} (l)\,\chi } \right)^2 } \right]^{\left| l \right|}  \simeq \rho ^{2\left| l \right|} \left( {1 - \frac{{2l\chi \xi _2 }}{{\rho ^2 }}} \right).
\]
From this equation follows that the \emph{nodal point} $W_l =0$ is shifted along $\xi_2$ on the distance ${\mathop{\rm sign}}\, l\,\chi$, while the \emph{center of gravity} of the vortex is shifted along the $\xi_2$ axis on distance $- l\chi$, i.e., in the opposite direction, see Fig.~\ref{Fig2} and cf. Ref.~\cite{OS}. Taking this into account, one can derive from Eq.~(\ref{14}) the differential equation describing the shift of the center of gravity of the vortex~(\ref{16}):
\begin{eqnarray}
\delta {\bf \dot r}_c ^{\{ \sigma ,l\} }  = \frac{{\sigma  + l}}{{k_c  }}\left( {{\bf \dot t} \times {\bf t}} \right),
\label{18}
\end{eqnarray}
where we substituted $\frac{{{\bm\nabla}_\bot \varepsilon _c }}{{2\varepsilon _c  }} = {\bf \dot t}$ from Eq.~(\ref{2}). Equation~(\ref{18}) represents correction to the ray Eqs.~(\ref{2}) determining motion of the center of gravity of a wave carrying well-defined spin and orbital AM. It is in agreement with the geometrical-optics predictions of the SHE~\cite{LZ,Bliokh1,OMN,DHH,NP} and OHE~\cite{Bliokh2}, which are directly related to the conservation of the total AM in the problem~\cite{OMN,Bliokh2}.

However, the $l$-dependent term in Eq.~(\ref{18}) is valid only until one can neglect perturbations in the beam shape caused by the diffraction in a gradient-index medium. In the lowest-order approximation, the diffraction-induced deformations are described by Eq.~(\ref{15}) with the right-hand-side terms neglected \cite{PS}. Typically, the beam acquires elliptical deformations at distances comparable with the characteristic inhomogeneity scale \cite{Ber}. These deformations do not affect the SHE, but they do affect the OHE, because elliptical deformations of a vortex beam dramatically change the intrinsic orbital AM carried by the beam \cite{Fedo} and the shift of the center of gravity of the vortex. In contrast to spin AM, the intrinsic orbital AM of the beam \emph{is not conserved} upon the diffraction in a gradient-index medium. Indeed, the operator of the orbital AM, $\hat{L} \propto -i {\bm \xi} \times {\bm\nabla}_\bot$, does not commute with the operator in the square brackets in the left-hand side of Eq.~(\ref{15}).

Equation (\ref{15}) cannot be solved analytically in the generic case even with the neglected right-hand side. Therefore, it is impossible to determine analytically the OHE shift of a diffracting beam in a gradient-index medium. However, Eq.~(\ref{15}) can be integrated numerically in each particular problem. Then, according to Eq.~(\ref{14}), the OHE shift and deformation of the beam can be taken into account by introducing an imaginary shift ${\bm \xi}\rightarrow {\bm \xi} -ik_c^{-1}{\bf \dot t}\, ds$ at each step $ds$. It should be noticed that the diffraction of the vortex beam also depends on the absolute value of the vortex charge, $|l|$, but, in contrast to the OHE, Eq.~(\ref{15}) is independent on the \emph{sign} of the vortex, ${\mathop{\rm sign}} (l)$.

To conclude, we have derived a parabolic-type equation which describes propagation and diffraction of paraxial beams in a gradient-index medium and accounts for SHE and OHE. Equations (\ref{13}) and (\ref{14}) enable one to separate the Hall effects, which lift the degeneracy of states with opposite helicities $\sigma  =  \pm 1$ and vorticities $l =  \pm \left| l \right|$, and the diffraction effects described by Eq.~(\ref{15}). While the SHE turns out to be diffraction-independent, the OHE is crucially affected by the beam deformations upon the diffraction in a gradient medium. Due to this, calculations of the OHE require numerical solution of the diffraction Eq.~(\ref{15}) in each particular problem.

This research was supported by the Linkage International Grant of the Australian Research Council.


\begin{thebibliography}{99}
\bibitem{remark1} {Differentiation of the wave field in the moving frame $\left( {{\bf e}_1 ,{\bf e}_2 ,{\bf t}} \right)$ produces additional terms which can be associated with the Coriolis effect: ${\bf \dot E} = \dot E_i {\bf e}_i  + \dot E_\parallel  {\bf t} + E_i {\bf \dot e}_i  + E_\parallel  {\bf \dot t}$. Indeed, the $\left( {{\bf e}_1 ,{\bf e}_2 ,{\bf t}} \right)$ frame rotates with the instant angular velocity ${\bf \Omega } = {\bf t} \times {\bf \dot t}$ with respect to $s$. The corresponding Coriolis term can be written as ${\bf \Omega } \times {\bf E} =  - \left( {{\bf E} \cdot {\bf \dot t}} \right){\bf t} + E_\parallel  {\bf \dot t} = E_i {\bf \dot e}_i  + E_\parallel  {\bf \dot t}$, where we used Eq. (4).}
\bibitem{remark2} Similarly to \cite{remark1}, the term $\frac{{\partial h}}{{\partial {\bm \xi }}}\frac{\partial }{{\partial {\bm \xi }}}$ in the Laplace operator can be associated with a Coriolis effect, because it is caused by derivatives of the frame parameters. Thus, both the SHE and OHE can be regarded as manifestations of the Coriolis effect under the bending of the wave trajectory.
\end{thebibliography}
\end{document}